\documentclass[a4paper,12pt]{article}

\usepackage{graphicx}

\pdfoutput=1

\usepackage{subcaption}
\usepackage[a4paper]{geometry}
\usepackage{graphicx}
\usepackage{amsmath}
\usepackage{amssymb}
\usepackage{float}
\usepackage{listings}
\usepackage{tikz}
\pdfoutput=1
\usepackage{epsfig,epsf}
 \usepackage{setspace}
 \usepackage{amsmath}
 \usepackage{array}
  \usepackage{tabu}
  \usepackage{graphicx}
\usepackage{subcaption}
\def\bea{\begin{eqnarray}}
\def\eea{\end{eqnarray}}

\begin{document}

\begin{center}
{\large\bf Reggeon webs, spin chains and the Odderon}

{Grigorios Chachamis}, {Agust{\' \i}n Sabio Vera$^\P$}

{Instituto de F{\' \i}sica Te{\' o}rica UAM/CSIC, Nicol{\'a}s Cabrera 15,\\ \& Universidad Aut{\' o}noma de Madrid, E-28049 Madrid, Spain.}

$^\P${E-mail: agustin.sabio@uam.es}
\end{center}

\centerline{\bf Abstract}
At high center-of-mass energies scattering amplitudes enjoy a hidden integrability. An important example in QCD is Odderon exchange, a composite state of three reggeized gluons, with can be understood as  a closed spin chain with periodic boundary conditions. 
In the $N=4$ supersymmetric Yang-Mills theory a similar open spin chain appears for the planar eight-point amplitude. We solve these two examples  in transverse momentum and rapidity space, introducing  the concept of weighted graph complexity and  its associated emerging scaling laws. 

\section{Introduction}

In the high energy regime of Quantum Chromodynamics (QCD) novel effective degrees of freedom, reggeized gluons, arise. They form composite states such as the Pomeron (2 reggeized gluons) and  Odderon~\cite{Odderon} (3 reggeized gluons), which are understood within the Balitsky-Fadin-Kuraev-Lipatov (BFKL) approach~\cite{BFKL}. Odderon exchange is described by solving the Bartels-Kwiecinski-Praszalowicz  (BKP) equation~\cite{BKP} which can be studied using results from integrable systems~\cite{Integrability,Recent} since it can be mapped to a closed spin chain (CSC)~\cite{Integrability}. In~\cite{Chachamis:2016ejm}, we solved the Odderon equation using Monte Carlo integration methods and, recently, also a similar problem associated to open spin chains~\cite{Chachamis:2018bys}. 

The solution to the BKP equation for 3 Reggeons gives a 6-point amplitude with off-shell external gluons  (written as $f_\omega \left({\bf p}_1,{\bf p}_2,{\bf p}_3 \right)$). ${\bf p}_i$ are 6 2-dimensional vectors such that ${\bf p}_{i=1,2,3}$ are at rapidity $Y$ and ${\bf p}_{i=4,5,6}$ at rapidity 0.  $\omega$ is the Mellin-conjugate variable of $Y$. Using a mass-dimension regulator $\lambda$ the gluon Regge trajectory is  (with ${\bar \alpha}_s = \alpha_s N_c / \pi$) $\omega ({\bf p}) = - \frac{{\bar \alpha}_s}{2} \ln{\frac{{\bf p}^2}{\lambda^2}}$. The $\lambda$ dependence is removed when the square of a Lipatov's emission vertex,  
\begin{eqnarray}
\xi \left({\bf p}_i,{\bf p}_j,{\bf p}_k,{\bf k}\right) ~= \frac{{\bar \alpha}_s}{4} \frac{ \theta({\bf k}^2 - \lambda^2)}{\pi {\bf k}^2}
 \Bigg(1+\frac{({\bf p}_i+{\bf k})^2 {\bf p}_j^2-({\bf p}_i+{\bf p}_j)^2 {\bf k}^2}{{\bf p}_i^2 ({\bf k}-{\bf p}_2)^2}\Bigg),
 \label{xifunction}
\end{eqnarray}
is integrated over ${\bf k}$.  $\xi$ couples 2 reggeized gluons of momentum ${\bf p}_i$ and ${\bf p}_j$ via the momentum ${\bf k}$. This corresponds to pairwise interactions in the associated CSC. The BKP equation for Odderon is
\begin{eqnarray}
\left(\omega - \omega({\bf p}_1) - \omega({\bf p}_2) - \omega({\bf p}_3) \right) 
 f_\omega \left({\bf p}_1,{\bf p}_2,{\bf p}_3\right) &=& \delta^{(2)}  \left({\bf p}_1-{\bf p}_4 \right) \delta^{(2)}  \left({\bf p}_2-{\bf p}_5 \right) 
 \delta^{(2)}  \left({\bf p}_3-{\bf p}_6 \right)  \nonumber\\
 &&\hspace{-8.5cm}+  \int d^2 {\bf k} \, \xi \left({\bf p}_1,{\bf p}_2,{\bf p}_3,{\bf k}\right)
 f_\omega \left({\bf p}_1+{\bf k},{\bf p}_2-{\bf k},{\bf p}_3\right) +  \int d^2 {\bf k} \,  \xi \left({\bf p}_2,{\bf p}_3,{\bf p}_1,{\bf k}\right)
  f_\omega \left({\bf p}_1,{\bf p}_2+{\bf k},{\bf p}_3-{\bf k} \right) \nonumber\\
 &&\hspace{-4cm}+  \int d^2 {\bf k}  \, \xi \left({\bf p}_1,{\bf p}_3,{\bf p}_2,{\bf k}\right)f_\omega \left({\bf p}_1+{\bf k},{\bf p}_2,{\bf p}_3-{\bf k}\right).
\label{BKPeq} 
\end{eqnarray}
In~\cite{Chachamis:2016ejm} we iterated the BKP ternary kernel acting on the initial condition.  We found that the gluon Green function  grows   for small values of $Y$ and then fastly decreases at higher $Y$. This is compatible with previous approaches~\cite{Bartels:1999yt}.  Now we review our method~\cite{Chachamis:2018bys} to solve an equation similar to Eq.~(\ref{BKPeq}) which corresponds to an integrable open spin chain (OSC)~\cite{Lipatov:2009nt}. 

\section{The open spin chain}

In the 8-gluon amplitude for the $N=4$ supersymmetric planar theory in certain physical regions the 3  Reggeized gluon exchange is important~\cite{Bartels:2008ce}. Its contribution to the amplitude corresponds to an OSC fulfilling the BKP-like integral equation:
\begin{eqnarray}
\left(\omega - \omega({\bf p}_1) - \omega({\bf p}_2) - \omega({\bf p}_3) \right) 
 f_\omega \left({\bf p}_1,{\bf p}_2,{\bf p}_3\right) &=&  \delta^{(2)}  \left({\bf p}_1-{\bf p}_4 \right) \delta^{(2)}  \left({\bf p}_2-{\bf p}_5 \right) 
 \delta^{(2)}  \left({\bf p}_3-{\bf p}_6 \right)  \nonumber\\
 &&\hspace{-8.9cm}+  \int d^2 {\bf k} \, \xi \left({\bf p}_1,{\bf p}_2,{\bf p}_3,{\bf k}\right)
 f_\omega \left({\bf p}_1+{\bf k},{\bf p}_2-{\bf k},{\bf p}_3\right) +  \int d^2 {\bf k} \,  \xi \left({\bf p}_2,{\bf p}_3,{\bf p}_1,{\bf k}\right)
  f_\omega \left({\bf p}_1,{\bf p}_2+{\bf k},{\bf p}_3-{\bf k} \right).
\label{OpenBKP} 
\end{eqnarray}
Using ${\cal O} ({\bf k}) \otimes f  \equiv \xi \left({\bf p}_1,{\bf p}_2,{\bf p}_3,{\bf k}\right)
 f \left({\bf p}_1+{\bf k},{\bf p}_2-{\bf k},{\bf p}_3\right)+ \xi \left({\bf p}_2,{\bf p}_3,{\bf p}_1,{\bf k}\right)
  f \left({\bf p}_1,{\bf p}_2+{\bf k},{\bf p}_3-{\bf k} \right)$: 
\begin{eqnarray}
\left(\omega - \omega({\bf p}_1) - \omega({\bf p}_2) - \omega({\bf p}_3) \right) 
 f_\omega \left({\bf p}_1,{\bf p}_2,{\bf p}_3\right) &=& \delta^{(2)}  \left({\bf p}_1-{\bf p}_4 \right) \delta^{(2)}  \left({\bf p}_2-{\bf p}_5 \right) 
 \delta^{(2)}  \left({\bf p}_3-{\bf p}_6 \right) \nonumber\\
&& \hspace{-4.cm} +\int d^2 {\bf k} \,  {\cal O} ({\bf k}) \otimes f_\omega \left({\bf p}_1,{\bf p}_2,{\bf p}_3\right).
\end{eqnarray}
Iterating and trading $\omega$  for a $Y$ we can write the solution in the form
\begin{eqnarray}
f_{\rm BKP}^{\rm adjoint} \left({\bf p}_1,{\bf p}_2,{\bf p}_3, Y\right)  &=&   e^{(\omega({\bf p}_1) + \omega({\bf p}_2) + \omega({\bf p}_3)) Y} \delta^{(2)}  \left({\bf p}_1-{\bf p}_4 \right) \delta^{(2)}  \left({\bf p}_2-{\bf p}_5 \right) \delta^{(2)}  \left({\bf p}_3-{\bf p}_6 \right)
\nonumber\\
 &&\hspace{-3cm} + \sum_{n=1}^\infty \Bigg\{\prod_{i=1}^n \int_0^{y_{i-1}} d y_i 
 \int d^2 {\bf k}_i e^{(\omega({\bf p}_1) + \omega({\bf p}_2) + \omega({\bf p}_3)) (y_{i-1}- y_i)} {\cal O} ({\bf k}_i) \otimes \Bigg\} \nonumber\\
 &&\hspace{-2.5cm}e^{(\omega({\bf p}_1) + \omega({\bf p}_2) + \omega({\bf p}_3)) y_n} 
 \delta^{(2)}  \left({\bf p}_1-{\bf p}_4 \right) \delta^{(2)}  \left({\bf p}_2-{\bf p}_5 \right) \delta^{(2)}  \left({\bf p}_3-{\bf p}_6 \right).
 \label{IterativeEqn}
\end{eqnarray}
In a similar fashion to~\cite{BFKLex1} we have shown~\cite{Chachamis:2018bys} that this Green function admits the factorization
\begin{eqnarray}
f_{\rm BKP}^{\rm adjoint} \left({\bf p}_1,{\bf p}_2,{\bf p}_3, {\rm Y}\right) &=& \left(\frac{\lambda^2}{\sqrt{{\bf p}_1^2 {\bf p}_3^2}}\right)^{\frac{\bar{\alpha}_s {\rm Y}}{2}} {\widehat f}_{\rm BKP}^{\, \, {\rm adjoint}} \left({\bf p}_1,{\bf p}_2,{\bf p}_3, {\rm Y}\right).
\end{eqnarray}
We investigate the infrared finite function ${\widehat f}_{\rm BKP}$ and evaluate the solution to the OSC and compare it to the CSC. We have 3 Reggeons (1), (2), (3) vertically aligned and ordered with (1) to the left and (3) to the right. In the CSC each Reggeon can interact with any of the other two via gluon {\it rungs}. A rung between (1) and (2) is denoted by L, between (2) and (3) by R and between (1) and (3) by M. In each Feynman diagram there are nodes with 2 or 3 attached legs. The  square {\it adjacency matrix} has as off-diagonal elements the number of lines connecting the vertices $i$ and $j$. 

\begin{figure}[H] 
    \centering
    \includegraphics[width=0.6\linewidth]{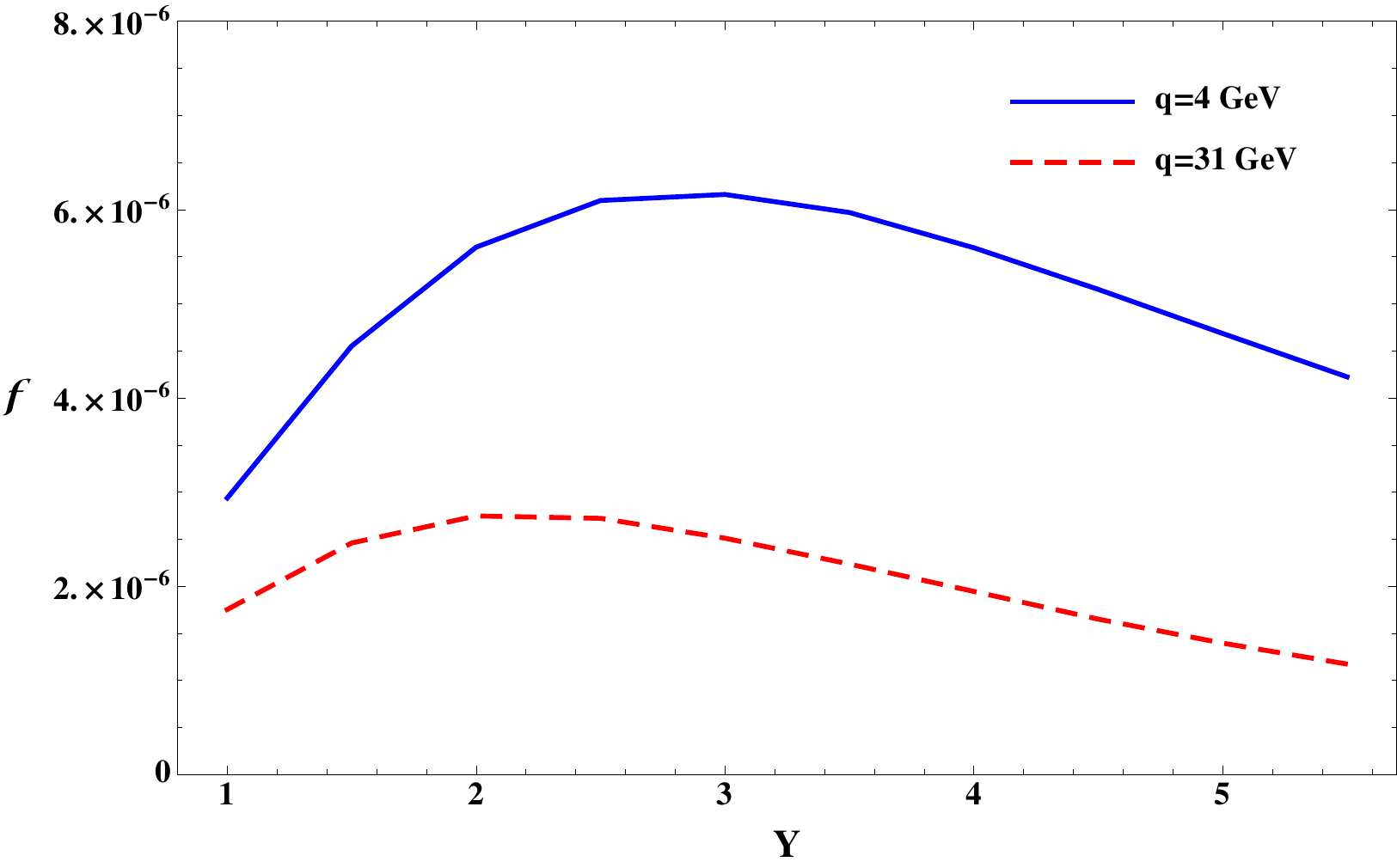} 
        \vspace{-.4cm}
 \caption{Energy dependence of $f$ for $q = 4, 31$ GeV for the closed spin chain.}
 \label{en1}
\end{figure}

\begin{figure}[H] 
    \centering
    \includegraphics[width=0.6\linewidth]{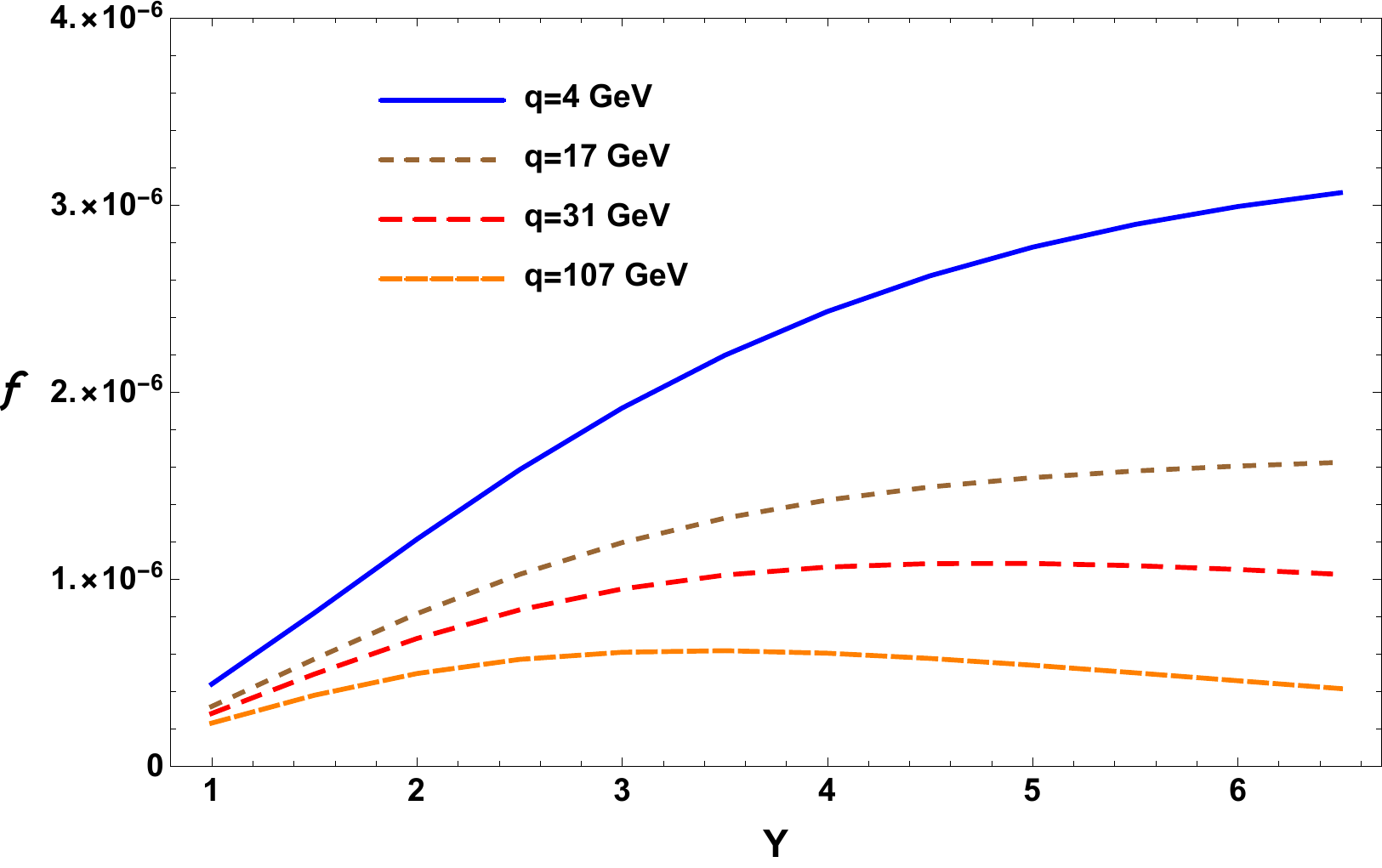} 
    \vspace{-.4cm}
 \caption{Energy dependence of $f$ for $q = 4, 17, 31, 107$ GeV  for the open spin chain. }
  \label{en2} 
\end{figure}

In~\cite{Chachamis:2016ejm}, we found the Odderon Green function with a method derived from our code {\tt BFKLex}~\cite{BFKLex1,BFKLex2}. Here we focus on an example of numerical external configurations which are given in polar coordinates (modulus in GeV, azimuthal angle) by ($Y$ ranges from 3.5 to 5.5)
\begin{align} 
& {\bf p}_1 = (10, 0) ;                             {\bf p}_2  = (20, \pi);  {\bf p}_3 = {\bf q}-{\bf p}_1-{\bf p}_2 ; \nonumber \\ 
& {\bf p}_4  = (20, 0);   {\bf p}_5 =   (25, \pi);   {\bf p}_6 = {\bf q}-{\bf p}_4 -{\bf p}_5.
\label{momenta}
 \end{align}
\begin{figure}[H]
\centering
 \includegraphics[width=0.6\linewidth]{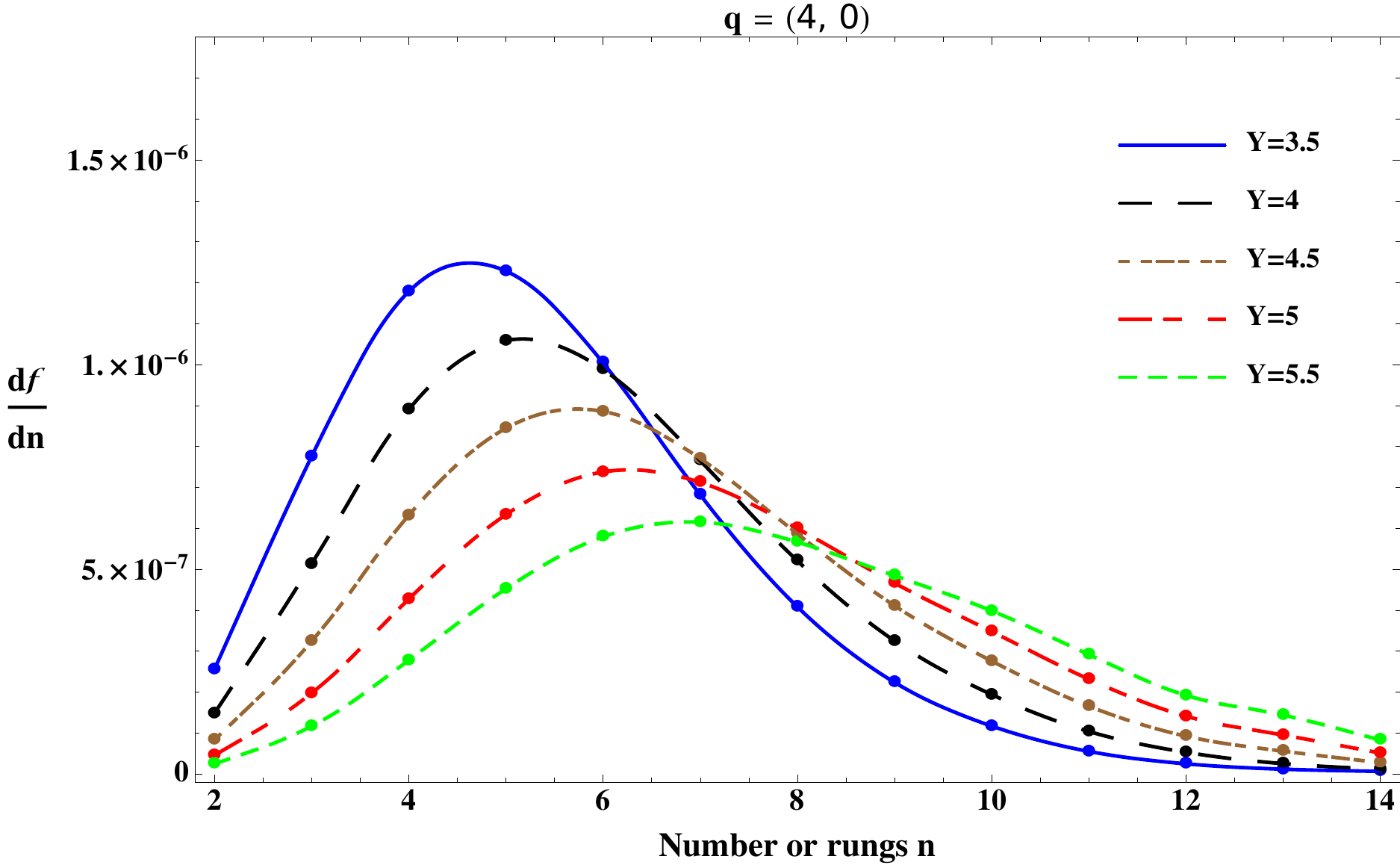} 
     \vspace{-.4cm}
\caption{{\it Multiplicity} plot for $q = 4$ GeV in the closed spin chain.}  
    \label{q4-1} 
\end{figure}
\begin{figure}[H]
\centering
    \includegraphics[width=0.6\linewidth]{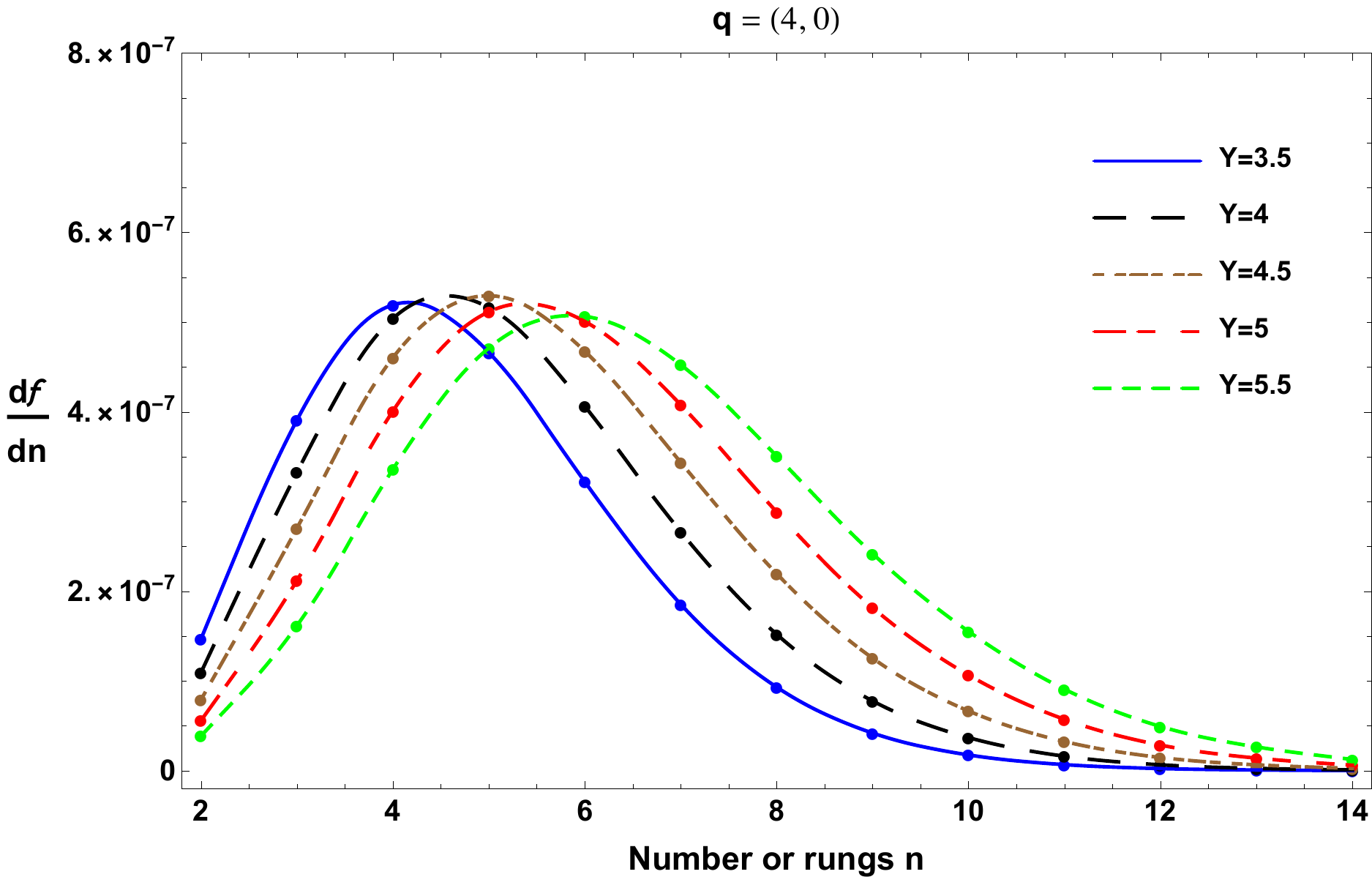}
        \vspace{-.4cm}
\caption{{\it Multiplicity} plot for $q = 4$ GeV in the open spin chain.}  
    \label{q4-2} 
\end{figure}
\begin{figure}
\centering
    \includegraphics[width=0.6\linewidth]{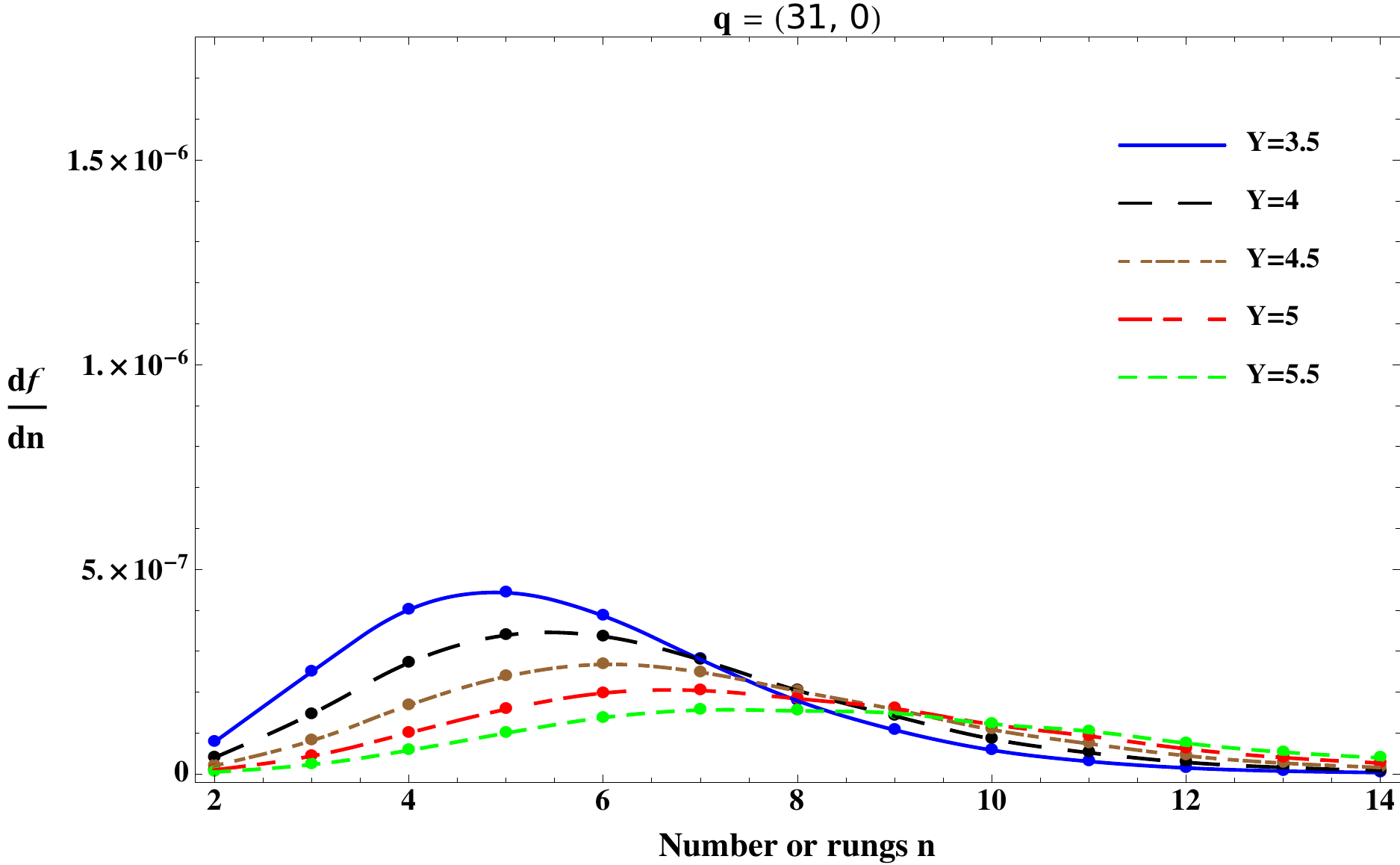}
        \vspace{-.4cm}
\caption{{\it Multiplicity} plot for $q = 31$ GeV in the closed spin chain.}  
\label{q31-1}
\end{figure}
\begin{figure}
\centering
    \includegraphics[width=0.6\linewidth]{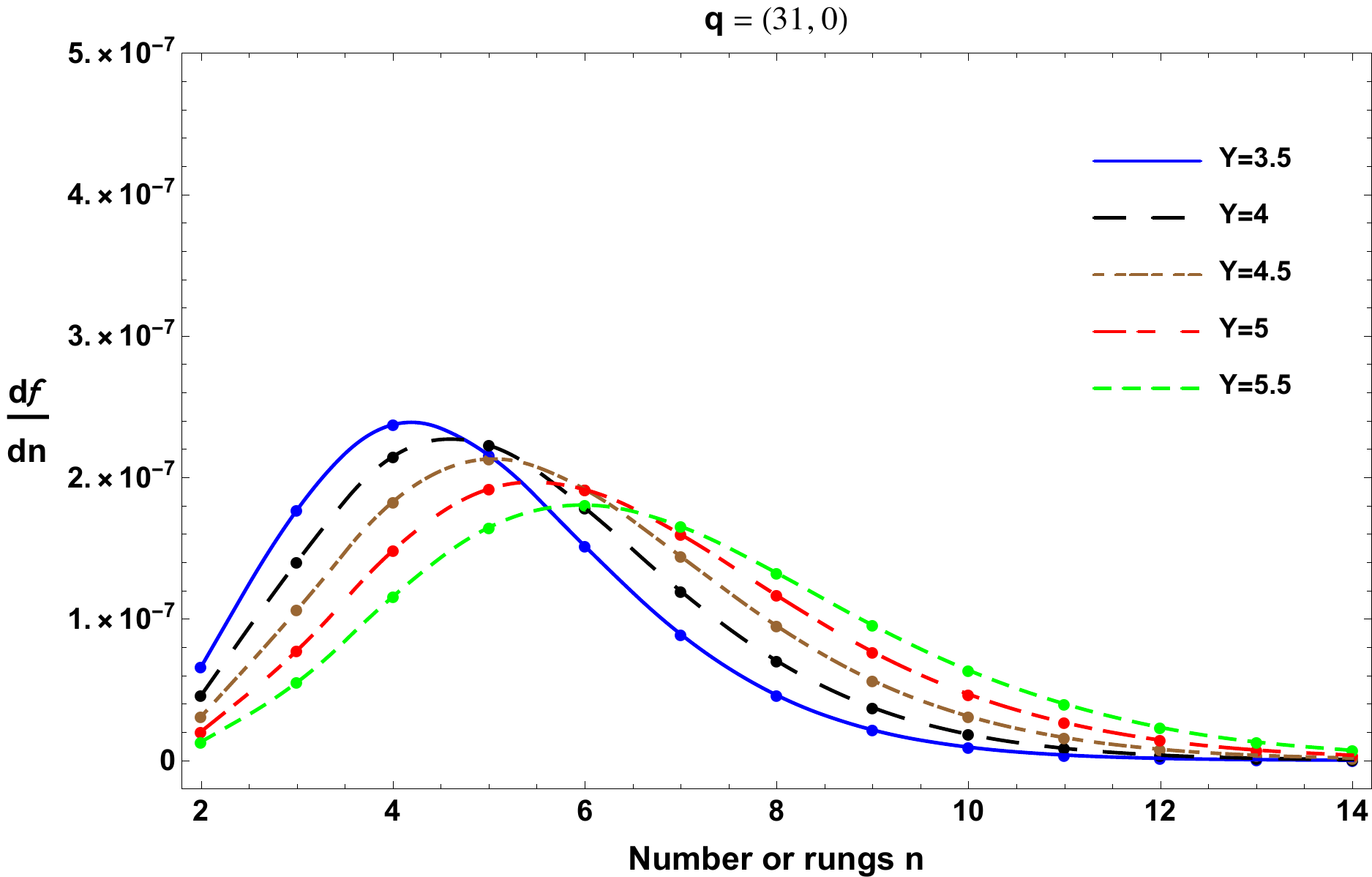}
        \vspace{-.4cm}
\caption{{\it Multiplicity} plot for $q = 31$ GeV in the open spin chain.}  
\label{q31-2}
\end{figure}
We show  {\it multiplicity} (number of rungs) plots with momentum transfer ${\bf q} = (4,0)$ (Fig.~\ref{q4-2}) and ${\bf q} = (31,0)$ (Fig.~\ref{q31-2}), and compare them to those in the CSC, Figs.~\ref{q4-1} and~\ref{q31-1}. They follow a Poisson-like distribution. For OSC  the decrease  with $Y$ in the maximal value happens very slowly when compared to CSC.  The 6-point gluon Green functions correspond to the areas under these distributions. Their change with $Y$ is in Fig.~\ref{en1}, \ref{en2}. The CSC  has its peak at low  $Y<3$ for ${\bf q} = (4, 0)$ and ${\bf q} = (31, 0)$ to then decrease very quickly. The OSC  for ${\bf q} = (4, 0)$ increases monotonically. For ${\bf q} = (17, 0)$ it rises more slowly. For ${\bf q} = (31, 0)$, $Y>6$, it decreases. ${\bf q} = (107, 0)$ shows a  maximum at $Y\sim3.4$.

\section{Weighted Graph Complexity}

{\it Spanning trees} are paths in a graph connecting all its nodes without any loops. 
Their total number for a graph ${\cal G}$ is its  {\it complexity} . The {\it degree} of a vertex $i$, $d_{\cal G} (i)$, is the number of half-edges in contact with it. Our graphs have degree 2 and 3. With $|V|$ being the number of vertices in ${\cal G}$ then we have that the {\it degree matrix} of ${\cal G}$, $D_{\cal G}$, is the $|V| \times |V|$ square matrix with diagonal elements $d_{\cal G} (i)$. Its matrix elements are $D_{\cal G} (i,j) = d_{\cal G} (i) \delta_{i j}$. Its  adjacency matrix  $A_{\cal G} $ is the $|V| \times |V|$ square matrix with off-diagonal elements being the number of propagators connecting $i$ with $j$. The {\it Laplacian matrix} is $L_{\cal G} = D_{\cal G} - A_{\cal G}$ and it is very useful to find the associated complexity of a Reggeon graph. For this we use the {\it Matrix-Tree theorem} (Kirchhoff, 1847) stating that the complexity of a graph is the determinant of any of its principal minors.  For a fixed number of rungs the number of CSC graphs is much larger than for the corresponding OSC where the latter are a subset of the former.
\begin{figure}[H]
\begin{center}
\begin{subfigure}{.55\textwidth}
\includegraphics[height=7cm]{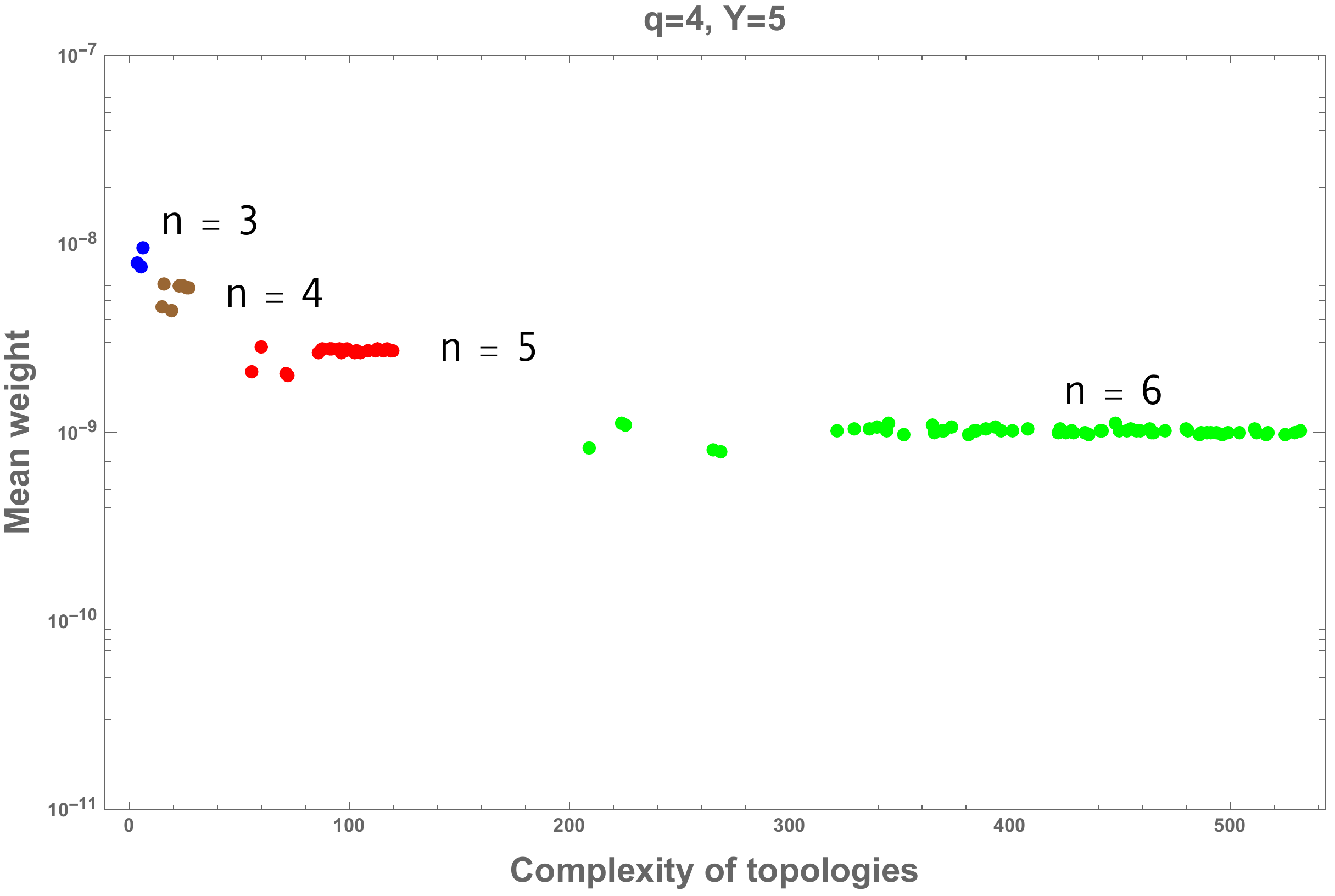}
\vspace{-.7cm}
\caption{Closed spin chain, $Y=3$.} \label{ClosedY5}
\end{subfigure}
\end{center}
\caption{Average weight per complexity class for closed Reggeon webs.}
\label{ClosedComplexityPlots}
\end{figure}
\begin{figure}[H]
\begin{center}
\begin{subfigure}{0.55\textwidth}
\includegraphics[height=7cm]{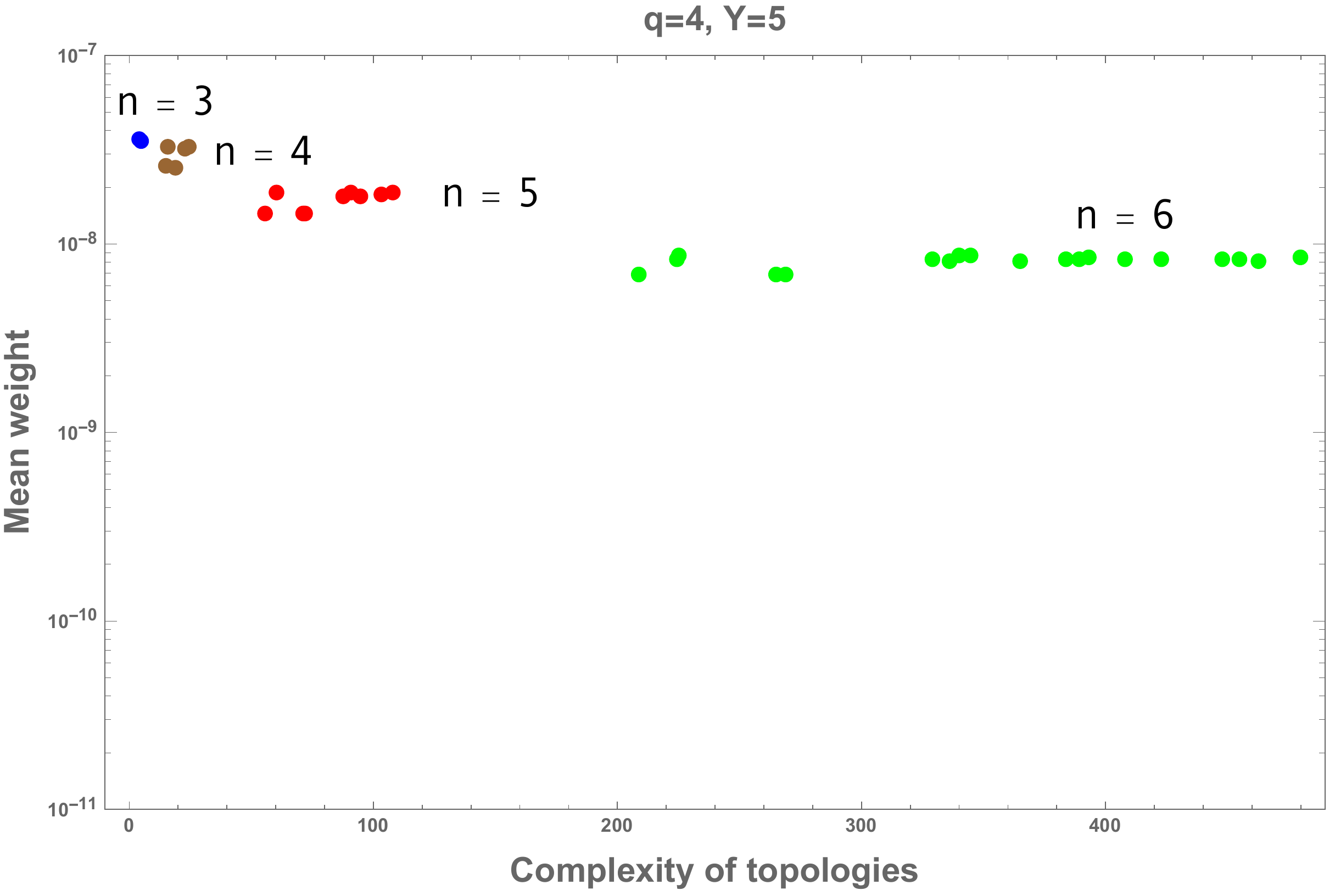}
\vspace{-.7cm}
\caption{Open spin chain, $Y=5$.} \label{ClosedY5}
\end{subfigure}
\end{center}
\vspace{-.7cm}
\caption{Average weight per complexity class for open Reggeon webs.}
\label{OpenComplexityPlots}
\end{figure}
We calculated the contribution to the gluon Green function from different graph complexities combined with their  statistical weight to our Monte Carlo solution to the BKP equation. The complexity of the graphs grows with the number of rungs $n$. We considered all those diagrams with the same complexity for a given $n$ and evaluated the average weight of their contribution to the gluon Green function, for a fixed value of the coupling and $Y$. 
We found a very interesting scaling behaviour shown as an example in Fig.~\ref{ClosedComplexityPlots} for the CSC and in Fig.~\ref{OpenComplexityPlots} for the OSC. For large values of complexity in each set with the same number of rungs all complexities contribute the same to the solution of the BKP equation both for close and open Reggeon graphs and values of the scattering energy.  We believe that this {\it complexity democracy} is linked to the underlying integrability found by Lipatov both in CSC and OSC.

\section{Conclusions}

The high energy limit of scattering amplitudes is very interesting. In QCD an integrable structure emerges which is related to a Heisenberg ferromagnet in a closed spin chain. We have solved this system using Monte Carlo integration. For the three reggeon case this is the  solution to the perturbative Odderon.  Lipatov found an integrable open spin chain in the evaluation of amplitudes in the $N=4$ supersymmetric Yang-Mills theory.  We have solved this open spin chain problem exactly. We have shown that the solution decreases with energy.  Our results can be extended to any number of reggeized gluons and higher order corrections.  We have found a very interesting scaling behavior in terms of a 
weighted complexity of the Feynman graphs contributing to the gluon Green function.  This  {\it complexity democracy}  is likely related to Lipatov's integrability.

\begin{flushleft}
{\bf \large Acknowledgements}
\end{flushleft}
We acknowledge support from the Spanish Government grants FPA2015-65480-P, FPA2016-78022-P and Spanish MINECO Centro de Excelencia Severo Ochoa Programme (SEV-2016-0597).

\end{document}